\documentclass[12pt]{article}

\usepackage{scicite}

\usepackage{times}
\usepackage{graphicx}
\usepackage{amssymb}
\usepackage{calrsfs,euscript,mathrsfs}
\usepackage{lineno}
\usepackage{deluxetable}

\newenvironment{myindentpar}[1]%
 {\begin{list}{}%
         {\setlength{\leftmargin}{#1}}%
         \item[]%
 }
 {\end{list}}

\newcommand\aj{\ref@jnl{AJ}}%
\newcommand\actaa{\ref@jnl{Acta Astron.}}%
\newcommand\araa{\ref@jnl{ARA\&A}}%
\newcommand\apj{\ref@jnl{ApJ}}%
\newcommand\apjl{\ref@jnl{ApJ}}%
\newcommand\apjs{\ref@jnl{ApJS}}%
\newcommand\ao{\ref@jnl{Appl.~Opt.}}%
\newcommand\apss{\ref@jnl{Ap\&SS}}%
\newcommand\aap{\ref@jnl{A\&A}}%
\newcommand\aapr{\ref@jnl{A\&A~Rev.}}%
\newcommand\aaps{\ref@jnl{A\&AS}}%
\newcommand\azh{\ref@jnl{AZh}}%
\newcommand\baas{\ref@jnl{BAAS}}%
\newcommand\caa{\ref@jnl{Chinese Astron. Astrophys.}}%
\newcommand\cjaa{\ref@jnl{Chinese J. Astron. Astrophys.}}%
\newcommand\icarus{\ref@jnl{Icarus}}%
\newcommand\jcap{\ref@jnl{J. Cosmology Astropart. Phys.}}%
\newcommand\jrasc{\ref@jnl{JRASC}}%
\newcommand\memras{\ref@jnl{MmRAS}}%
\newcommand\mnras{\ref@jnl{MNRAS}}%
\newcommand\na{\ref@jnl{New A}}%
\newcommand\nar{\ref@jnl{New A Rev.}}%
\newcommand\pra{\ref@jnl{Phys.~Rev.~A}}%
\newcommand\prb{\ref@jnl{Phys.~Rev.~B}}%
\newcommand\prc{\ref@jnl{Phys.~Rev.~C}}%
\newcommand\prd{\ref@jnl{Phys.~Rev.~D}}%
\newcommand\pre{\ref@jnl{Phys.~Rev.~E}}%
\newcommand\prl{\ref@jnl{Phys.~Rev.~Lett.}}%
\newcommand\pasa{\ref@jnl{PASA}}%
\newcommand\pasp{\ref@jnl{PASP}}%
\newcommand\pasj{\ref@jnl{PASJ}}%
\newcommand\qjras{\ref@jnl{QJRAS}}%
\newcommand\rmxaa{\ref@jnl{Rev. Mexicana Astron. Astrofis.}}%
\newcommand\skytel{\ref@jnl{S\&T}}%
\newcommand\solphys{\ref@jnl{Sol.~Phys.}}%
\newcommand\sovast{\ref@jnl{Soviet~Ast.}}%
\newcommand\ssr{\ref@jnl{Space~Sci.~Rev.}}%
\newcommand\zap{\ref@jnl{ZAp}}%
\newcommand\nat{\ref@jnl{Nature}}%
\newcommand\iaucirc{\ref@jnl{IAU~Circ.}}%
\newcommand\aplett{\ref@jnl{Astrophys.~Lett.}}%
\newcommand\apspr{\ref@jnl{Astrophys.~Space~Phys.~Res.}}%
\newcommand\bain{\ref@jnl{Bull.~Astron.~Inst.~Netherlands}}%
\newcommand\fcp{\ref@jnl{Fund.~Cosmic~Phys.}}%
\newcommand\gca{\ref@jnl{Geochim.~Cosmochim.~Acta}}%
\newcommand\grl{\ref@jnl{Geophys.~Res.~Lett.}}%
\newcommand\jcp{\ref@jnl{J.~Chem.~Phys.}}%
\newcommand\jgr{\ref@jnl{J.~Geophys.~Res.}}%
\newcommand\jqsrt{\ref@jnl{J.~Quant.~Spec.~Radiat.~Transf.}}%
\newcommand\memsai{\ref@jnl{Mem.~Soc.~Astron.~Italiana}}%
\newcommand\nphysa{\ref@jnl{Nucl.~Phys.~A}}%
\newcommand\physrep{\ref@jnl{Phys.~Rep.}}%
\newcommand\physscr{\ref@jnl{Phys.~Scr}}%
\newcommand\planss{\ref@jnl{Planet.~Space~Sci.}}%
\newcommand\procspie{\ref@jnl{Proc.~SPIE}}%

\newenvironment{sciabstract}{%
\begin{quote} \bf}
{\end{quote}}

\newcounter{lastnote}

\title{The Imprint of The Extragalactic Background Light
 in the Gamma-Ray Spectra of Blazars}

\date{}

\begin{document} 


%
%
%
%
%
%
%
%
%

\maketitle

\noindent
M.~Ackermann$^{1}$, 
M.~Ajello$^{2,3}\dagger$, 
A.~Allafort$^{2}$, 
P.~Schady$^{4}$, 
L.~Baldini$^{5}$, 
J.~Ballet$^{6}$, 
G.~Barbiellini$^{7,8}$, 
D.~Bastieri$^{9,10}$, 
R.~Bellazzini$^{11}$, 
R.~D.~Blandford$^{2}$
% ...
\medskip
\begin{enumerate}
\item[1.] Deutsches Elektronen Synchrotron DESY, D-15738 Zeuthen, Germany
\item[2.] W. W. Hansen Experimental Physics Laboratory, Kavli Institute for Particle Astrophysics and Cosmology, Department of Physics and SLAC National Accelerator Laboratory, Stanford University, Stanford, CA 94305, USA
\item[3.] Space Sciences Laboratory, 7 Gauss Way, University of California, Berkeley, CA 94720-7450, USA
\item[4.] Max-Planck Institut f\"ur extraterrestrische Physik, 85748 Garching, Germany
\item[5.] Universit\`a  di Pisa and Istituto Nazionale di Fisica Nucleare, Sezione di Pisa I-56127 Pisa, Italy
\item[6.] Laboratoire AIM, CEA-IRFU/CNRS/Universit\'e Paris Diderot, Service d'Astrophysique, CEA Saclay, 91191 Gif sur Yvette, France
\item[7.] Istituto Nazionale di Fisica Nucleare, Sezione di Trieste, I-34127 Trieste, Italy
\item[8.] Dipartimento di Fisica, Universit\`a di Trieste, I-34127 Trieste, Italy
\item[9.] Istituto Nazionale di Fisica Nucleare, Sezione di Padova, I-35131 Padova, Italy
\item[10.] Dipartimento di Fisica e Astronomia "G. Galilei", Universit\`a di Padova, I-35131 Padova, Italy
\item[11.] Istituto Nazionale di Fisica Nucleare, Sezione di Pisa, I-56127 Pisa, Italy
\item[12.] Santa Cruz Institute for Particle Physics, Department of Physics and Department of Astronomy and Astrophysics, University of California at Santa Cruz, Santa Cruz, CA 95064, USA
\item[13.] Dipartimento di Fisica ``M. Merlin" dell'Universit\`a e del Politecnico di Bari, I-70126 Bari, Italy
\item[14.] Istituto Nazionale di Fisica Nucleare, Sezione di Bari, 70126 Bari, Italy
\item[15.] Laboratoire Leprince-Ringuet, \'Ecole polytechnique, CNRS/IN2P3, Palaiseau, France
\item[16.] Institut de Ci\`encies de l'Espai (IEEE-CSIC), Campus UAB, 08193 Barcelona, Spain
\item[17.] INAF-Istituto di Astrofisica Spaziale e Fisica Cosmica, I-20133 Milano, Italy
\item[18.] Agenzia Spaziale Italiana (ASI) Science Data Center, I-00044 Frascati (Roma), Italy
\item[19.] Istituto Nazionale di Fisica Nucleare, Sezione di Perugia, I-06123 Perugia, Italy
\item[20.] Dipartimento di Fisica, Universit\`a degli Studi di Perugia, I-06123 Perugia, Italy
\item[21.] Center for Earth Observing and Space Research, College of Science, George Mason University, Fairfax, VA 22030, resident at Naval Research Laboratory, Washington, DC 20375, USA
\item[22.] National Research Council Research Associate, National Academy of Sciences, Washington, DC 20001, resident at Naval Research Laboratory, Washington, DC 20375, USA
\item[23.] INFN and Dipartimento di Fisica e Astronomia "G. Galilei", Universit\`a di Padova, I-35131 Padova, Italy, 
\item[24.] ASI Science Data Center, I-00044 Frascati (Roma), Italy
\item[25.] Laboratoire Univers et Particules de Montpellier, Universit\'e Montpellier 2, CNRS/IN2P3, Montpellier, France
\item[26.] Department of Physics, Stockholm University, AlbaNova, SE-106 91 Stockholm, Sweden
\item[27.] The Oskar Klein Centre for Cosmoparticle Physics, AlbaNova, SE-106 91 Stockholm, Sweden
\item[28.] Royal Swedish Academy of Sciences Research Fellow, funded by a grant from the K. A. Wallenberg Foundation
\item[29.] IASF Palermo, 90146 Palermo, Italy
\item[30.] INAF-Istituto di Astrofisica Spaziale e Fisica Cosmica, I-00133 Roma, Italy
\item[31.] Space Science Division, Naval Research Laboratory, Washington, DC 20375-5352, USA
\item[32.] Department of Physical Sciences, Hiroshima University, Higashi-Hiroshima, Hiroshima 739-8526, Japan
\item[33.] NASA Goddard Space Flight Center, Greenbelt, MD 20771, USA
\item[34.] INAF Istituto di Radioastronomia, 40129 Bologna, Italy
\item[35.] Department of Astronomy, Graduate School of Science, Kyoto University, Sakyo-ku, Kyoto 606-8502, Japan
\item[36.] Department of Physics, Royal Institute of Technology (KTH), AlbaNova, SE-106 91 Stockholm, Sweden
\item[37.] Research Institute for Science and Engineering, Waseda University, 3-4-1, Okubo, Shinjuku, Tokyo 169-8555, Japan
\item[38.] CNRS, IRAP, F-31028 Toulouse cedex 4, France
\item[39.] GAHEC, Universit\'e de Toulouse, UPS-OMP, IRAP, Toulouse, France
\item[40.] Department of Astronomy, Stockholm University, SE-106 91 Stockholm, Sweden
\item[41.] Istituto Nazionale di Fisica Nucleare, Sezione di Torino, I-10125 Torino, Italy
\item[42.] Department of Physics and Department of Astronomy, University of Maryland, College Park, MD 20742, USA
\item[43.] Hiroshima Astrophysical Science Center, Hiroshima University, Higashi-Hiroshima, Hiroshima 739-8526, Japan
\item[44.] Istituto Nazionale di Fisica Nucleare, Sezione di Roma ``Tor Vergata", I-00133 Roma, Italy
\item[45.] INTEGRAL Science Data Centre, CH-1290 Versoix, Switzerland
\item[46.] Institute of Space and Astronautical Science, JAXA, 3-1-1 Yoshinodai, Chuo-ku, Sagamihara, Kanagawa 252-5210, Japan
\item[47.] Department of Physics and Astronomy, University of Denver, Denver, CO 80208, USA
\item[48.] Max-Planck-Institut f\"ur Physik, D-80805 M\"unchen, Germany
\item[49.] Department of Physics and Center for Space Sciences and Technology, University of Maryland Baltimore County, Baltimore, MD 21250, USA
\item[50.] Center for Research and Exploration in Space Science and Technology (CRESST) and NASA Goddard Space Flight Center, Greenbelt, MD 20771, USA
\item[51.] Harvard-Smithsonian Center for Astrophysics, Cambridge, MA 02138, USA
\item[52.] Institut f\"ur Astro- und Teilchenphysik and Institut f\"ur Theoretische Physik, Leopold-Franzens-Universit\"at Innsbruck, A-6020 Innsbruck, Austria
\item[53.] Department of Physics, California Polytechnic State University, San Luis Obispo, CA 93401, USA
\item[54.] Department of Physics, University of Washington, Seattle, WA 98195-1560, USA
\item[55.] Max-Planck-Institut f\"ur Kernphysik, D-69029 Heidelberg, Germany
\item[56.] Space Sciences Division, NASA Ames Research Center, Moffett Field, CA 94035-1000, USA
\item[57.] NYCB Real-Time Computing Inc., Lattingtown, NY 11560-1025, USA
\item[58.] Astronomical Observatory, Jagiellonian University, 30-244 Krak\'ow, Poland
\item[59.] Department of Chemistry and Physics, Purdue University Calumet, Hammond, IN 46323-2094, USA
\item[60.] Instituci\'o Catalana de Recerca i Estudis Avan\c{c}ats (ICREA), Barcelona, Spain
\item[61.] NASA Postdoctoral Program Fellow, USA
\item[62.] Consorzio Interuniversitario per la Fisica Spaziale (CIFS), I-10133 Torino, Italy
\item[63.] Dipartimento di Fisica, Universit\`a di Roma ``Tor Vergata", I-00133 Roma, Italy
\item[64.] Department of Physics, Center for Cosmology and Astro-Particle Physics, The Ohio State University, Columbus, OH 43210, USA

\item[$\dagger$] majello@slac.stanford.edu, *buehler@stanford.edu, $\ddagger$ anita.reimer@uibk.ac.at

\end{enumerate}

\begin{sciabstract}
The light emitted by stars and  accreting compact objects
through the history  of the Universe
is encoded in the intensity of the extragalactic background light (EBL). 
Knowledge of the EBL is important to understand the nature of star formation and
galaxy evolution, but
direct measurements of the EBL are limited by 
Galactic and other foreground emissions. Here we report  an absorption
feature seen in the combined spectra of a sample of
gamma-ray blazars out  to a redshift of z$\sim$1.6.
This feature is caused by attenuation of gamma rays by the EBL at optical 
to UV frequencies,
and allowed us to measure  the EBL flux density in this frequency band.
\end{sciabstract}

The bulk of the intergalactic gas  in the Universe must have
been reionized between the epoch of cosmic recombination, when
the Universe was only 300,000 years old (z$\sim$1100),
and 1 billion years later
 (z$\sim$6) as indicated observationally by the spectra
of distant quasi-stellar objects \cite{fan06}.
However, the sources, modes and nature
of this cosmic reionization are largely unknown because most of this redshift
range has yet to be   explored.
Photoionization by UV radiation, produced  by the first stars and galaxies
of the Universe, represents the primary suspect for the ionizing process 
\cite{gilmore09,inoues10}.
Direct detection of the UV radiation fields 
is thus of fundamental importance, but at present extremely 
difficult \cite{inoues10}. 

An indirect but powerful means of probing the diffuse radiation fields
is through $\gamma$-$\gamma$ absorption of high-energy gamma rays
\cite{gould66,fazio70,stecker92}. In this process, a gamma-ray photon
of energy $E_{\gamma}$ and an EBL photon of energy $E_{EBL}$ annihilate
and create an electron-positron pair. 
This process occurs for head-on collisions when (e.g.)
 $E_{\gamma} \times E_{EBL}\geq 2(m_e c^2)^2$, where
$m_e c^2$ is the rest mass energy of the electron. This  introduces
an attenuation in the spectra of gamma-ray sources above a critical 
gamma-ray energy  of $E_{crit}(z)\approx  170(1+z)^{-2.38}$\,GeV
 \cite{franceschini08,note1}.

The detection of the gamma-ray horizon (i.e. the point beyond
which the emission of gamma-ray sources is strongly attenuated)
is one of the primary scientific drivers of the {\it Fermi}  Gamma-ray Space Telescope \cite{hartmann07,stecker07,kashlinsky07}.
Several attempts have been made in the past but none  detected
the long-sought EBL attenuation \cite{mannheim96,lat_ebl10,raue10}.
So far, limits  on the EBL density have been
inferred from the absence of absorption features in the spectra of individual
blazars \cite{aharonian06,lat_ebl10}, distant galaxies with bright gamma-ray emission powered by matter accreting onto central, massive black holes. 
While this feature is indeed difficult to
constrain for a single source, we show that
it is detected collectively  in the gamma-ray spectra of a sample of blazars
as a cut-off that changes amplitude and energy with redshift.
We searched for an attenuation of the spectra of blazars in the 1--500\,GeV band
using the first 46 months of observations of the Large Area Telescope (LAT) 
on board the {\it Fermi} satellite.
At these energies gamma rays are absorbed by EBL photons in the optical to UV range.  
Thanks to the large energy and redshift coverage, {\it Fermi}-LAT measures
the intrinsic (i.e. unabsorbed)  spectrum  up to $\sim$100\,GeV for any blazar at z$<$0.2, and up to $\sim$15\,GeV for any redshift.

%
%
%
The LAT has detected $>$1000 blazars to date \cite{2LAC}. 
We restricted our search to a 
subset of 150 blazars of the BL Lacertae (BL Lac) type
that are significantly detected above 3\,GeV,
because of the expected lack of intrinsic absorption \cite{reimer07}.
 The sample covers a redshift range 0.03--1.6 \cite{som,rau12}.
The critical energy is therefore always $\geq$25\,GeV,
which means that the spectrum measured below this energy is unabsorbed
and a true representation of the intrinsic spectrum of the source.
We thus determined the intrinsic source spectrum
relying on data between 1\,GeV and the critical energy E$_{crit}$
and extrapolated it to higher energies.
By combining all the spectra we were able to determine,  
the average deviation, above the critical energy,
of the measured spectra from the intrinsic ones,
which ultimately 
provides a measurement of the optical depth $\tau_{\gamma\gamma}$.

The analysis was performed using the {\it Fermi} Science Tools \cite{note2}.
We determined the spectral parameters of each blazar
by maximizing the likelihood of a given source model. 
The model comprised the Galactic and isotropic diffuse components 
and all sources in the second {\it Fermi} LAT catalog \cite{2FGL}
within a region of interest (ROI) of 15$^{\circ}$ radius. 
We modeled the spectra
of the sources in our sample
as parabolic in the logarithmic space of energy and flux
(see Eq.~2 in \cite{2FGL} for a definition). 
Their spectra were modified  by a term
$e^{-\tau_{\gamma\gamma}(E,z)}$ that describes the absorption
of gamma-ray photons on the EBL. In the above we defined
$\tau_{\gamma\gamma}(E,z) = b \cdot \tau^{model}_{\gamma\gamma}(E,z)$,
where the $\tau^{model}_{\gamma\gamma}(E,z)$ is the optical depth predicted
by EBL models \cite{kneiske04,stecker06,franceschini08,finke10,dominguez11}
and $b$ is a scaling variable, left free in the likelihood maximization.
In particular, this allowed us to assess the likelihood of 
two important scenarios:
i) there is no EBL attenuation ($b$=0), 
ii) the model prediction is correct ($b$=1).

\begin{figure*}[ht!]
  \begin{center}
  \begin{tabular}{c}
    \includegraphics[scale=0.9]{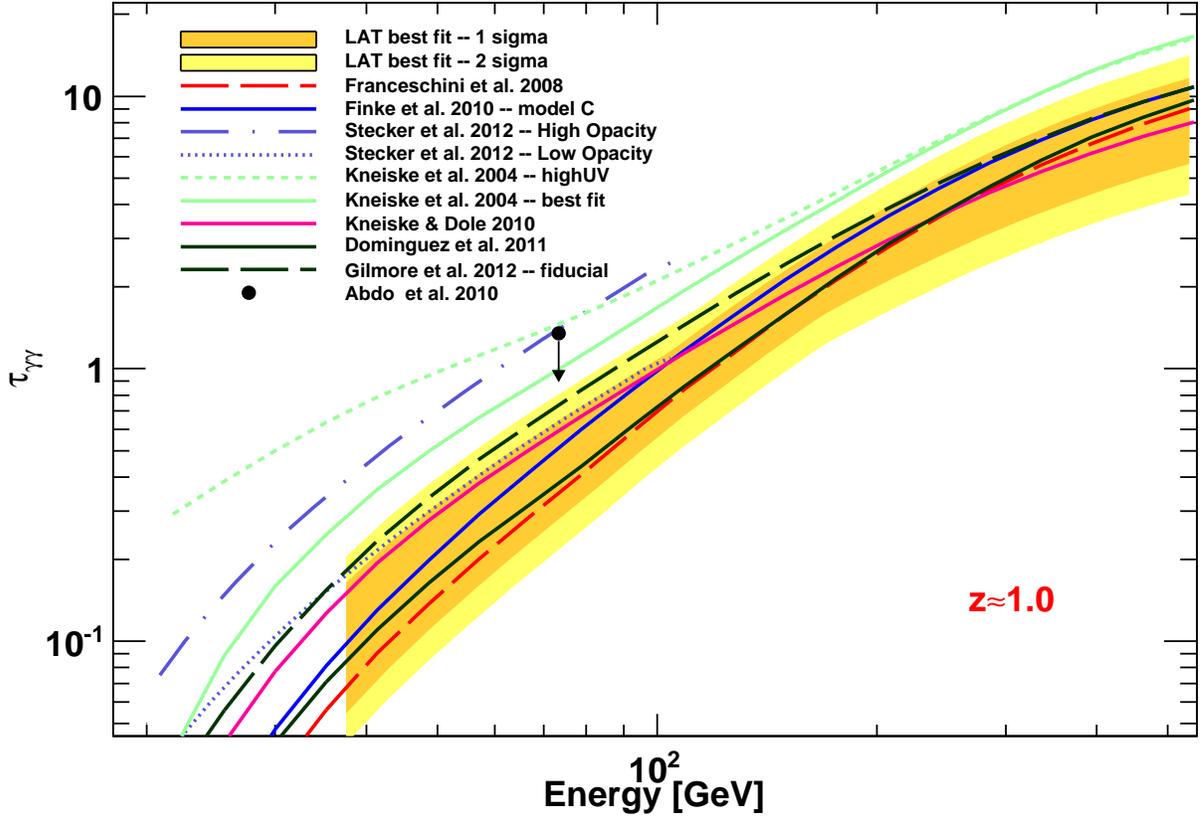} 
\end{tabular}
  \end{center}
  \caption{Measurement, at the 68\,\% and 95\,\% confidence levels (including
systematic uncertainties added in quadrature),
 of the opacity  $\tau_{\gamma\gamma}$  from the best fits  to the {\it Fermi} data
compared to predictions of
EBL models.  The plot shows the measurement at z$\approx$1 which is the 
average redshift of the most constraining redshift interval (i.e. 
0.5$\leq$z$<$1.6).
The {\it Fermi}-LAT measurement was derived combining the limits
on the best-fit EBL models.
The downward arrow represents the 95\,\% upper limit
on the opacity at z=1.05 derived in \cite{lat_ebl10}. For
clarity this figure shows only a selection of the models we
tested while the full list is reported in Table~S1.
The EBL models of \cite{stecker12}, which are not
defined for E$\geq$250$/(1+z)$\,GeV and thus could not be used, are
reported here for completeness.
\label{fig:tau}}
\end{figure*}

We combined the data from all the ROIs in a global fit that determined
the common parameter $b$ for a given EBL model
(see Table~S1). 
All those models  with a minimal EBL density based on (or compatible with)
 resolved galaxy counts 
\cite{franceschini08,gilmore09,kneiske10,finke10,dominguez11,gilmore12}
were found to be  acceptable descriptions of the {\it Fermi} data
 (i.e are consistent with $b$=1 within $\approx$25\,\%, see also Figure~\ref{fig:tau}) yielding
a significance of the absorption feature of up to $\sim$6\,$\sigma$.
Models  that predict a larger intensity of the EBL particularly in the UV \cite{kneiske04,stecker06}
would
produce a stronger-than-observed attenuation feature and are therefore 
incompatible with the {\it Fermi} observations.
Our measurement points to a minimal level of the optical-UV EBL up to redshift z$\approx$1.6 which combined
with the upper limits \cite{aharonian06,mazin07,magic08}
derived at lower redshift (using observations of blazars at TeV energies) 
on the near-infrared EBL highlights the conclusion
that most of the EBL intensity
can  be explained by the measured galaxy emission.

\begin{figure*}[ht!]
  \begin{center}
  \begin{tabular}{c}
    \includegraphics[scale=0.63]{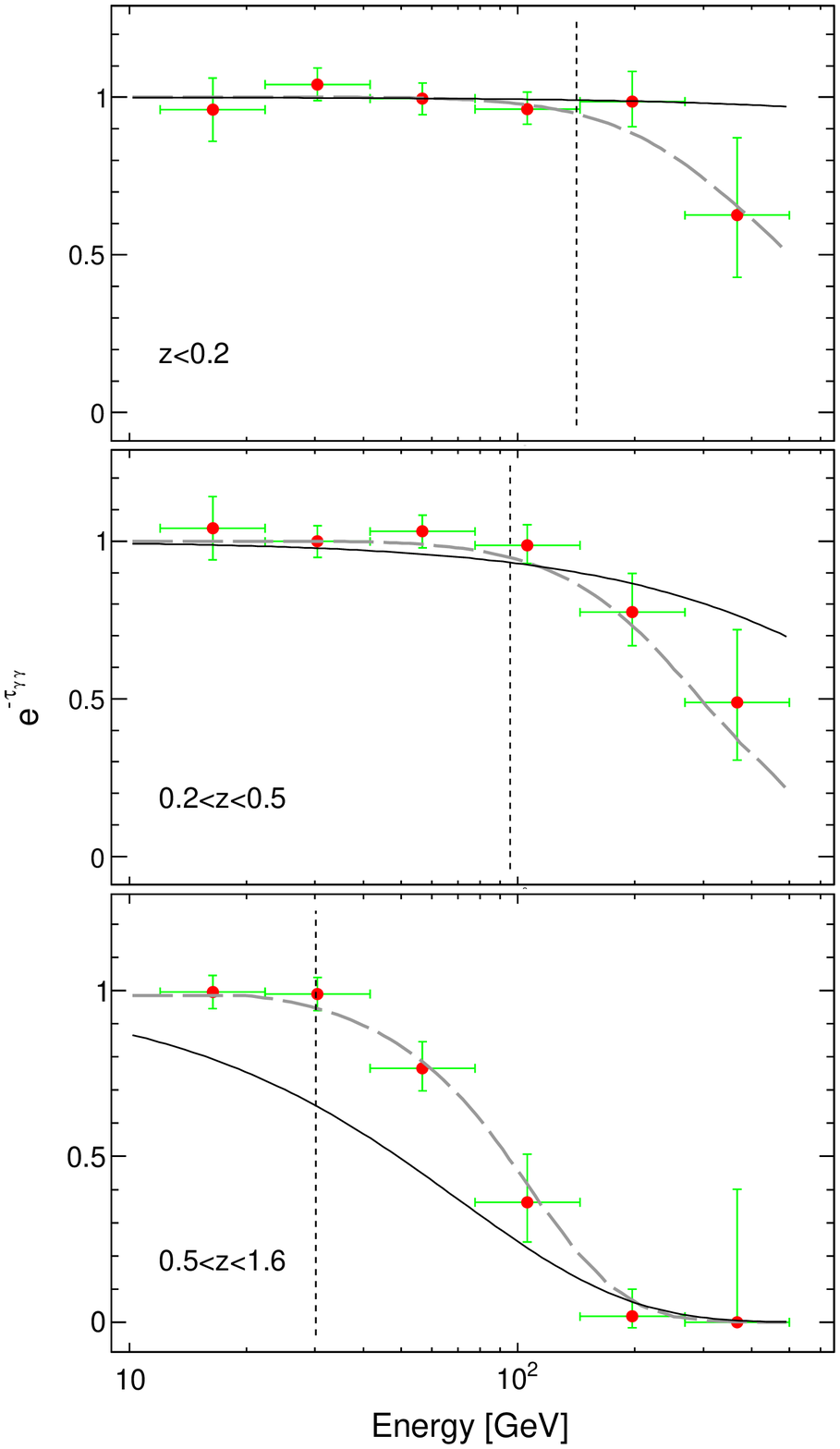} 
\end{tabular}
  \end{center}
  \caption{Absorption feature present in the spectra of BL Lacertae
objects as a function of increasing redshift (data points, from top to bottom).
The dashed curves show the attenuation expected for the sample of sources
by averaging, in each redshift and energy bin, the opacities of the sample
(the model of \cite{franceschini08} was used)
and multiplying this average by the best-fit scaling parameter $b$ 
obtained independently  in each redshift interval.
The vertical
line shows the critical energy E$_{crit}$ 
below which $\leq$5\,\% of the source photons
are absorbed by the EBL. The thin solid curve represents the best-fit model
assuming that all the sources have an intrinsic exponential cut-off
and that blazars follow the blazar sequence model of \cite{fossati99,ghisellini09b}.
\label{fig:residuals}}
\end{figure*}

Our measurement relies on the accuracy of the extrapolation
of the intrinsic spectra of the sources
above the critical energy \cite{note3}.
This in turn depends on a precise description of the gamma-ray spectra
by our source parametrization.
To verify that this is the case and 
to exclude the possibility that the detected absorption feature is  intrinsic
to the gamma-ray sources\cite{reimer07}, 
we performed the analysis in 3 independent
redshift intervals (z$<$0.2, 0.2$\leq$z$<$0.5, and 0.5$\leq$z$<$1.6).
The deviations from the intrinsic spectra
in the three redshift intervals are  displayed in Figure~\ref{fig:residuals}.
In the local Universe (z$<$0.2), EBL absorption is negligible  in most of
the  
{\it Fermi}-LAT energy band (E$_{crit}\geq$120\,GeV). The 
lowest redshift interval therefore reveals directly the intrinsic 
spectra of the sources and shows that our spectral parametrization is 
accurate \cite{som}.
The absorption feature is clearly visible above the critical energy in 
the higher redshift bins. Its amplitude and modulation in energy 
evolve with redshift as expected for EBL absorption. 
In principle, the observed attenuation could be due
to a spectral cut-off that is intrinsic to the gamma-ray sources.
The absence of a cut-off in the spectra of sources with z$<$0.2 would require
that the properties of BL Lacs change with redshift or luminosity.
It remains an issue of debate whether such evolution exists 
\cite{ghisellini98,fossati99,ghisellini09b,meyer11}.
However, in  case it were present, the intrinsic cut-off would be
expected  to evolve differently with redshift than we observe.
To illustrate this effect, we fitted the blazar sample
assuming that all the sources have an exponential cut-off at an energy $E_0$. 
From source to source the observed cut-off energy changes
because of the source redshift and because we assumed that 
blazars as a population are distributed in a sequence such as that
proposed in \cite{ghisellini98,fossati99,ghisellini09b,meyer11}.
$E_0$ was fitted to the data globally like $b$ above.
As apparent from Figure~\ref{fig:residuals}, it appears  difficult 
to reconcile the observed feature
with an intrinsic characteristic of the blazars' spectra.
We therefore associate the spectral feature to the  EBL absorption.

At energies  $\leq$100\,GeV, gamma rays observed at Earth and
coming from  redshift $\geq$1  interact mostly with  UV photons of
$\geq$5\,electron volts.  An UV background in excess of the light emitted 
by resolved galaxies can be produced locally by AGN or at higher redshift
(z$\approx$7-15) by low-metallicity massive stars \cite{santos02}. 
By comparing the results from the best-fit EBL models,
we measured the UV component of the EBL to have an intensity
of 3($\pm1$)\,nW m$^{-2}$ sr$^{-1}$ at z$\approx$1. 
A contribution to the UV background from AGN as large as the
one predicted by \cite{haardt96} (i.e., $\approx10$\,nW m$^{-2}$ sr$^{-1}$)
and used in the EBL model  of \cite{kneiske04} is thus excluded
by our analysis at high confidence. However, 
the recent prediction \cite{haardt12} of the UV background from AGN 
 ($\approx2$\,nW m$^{-2}$ sr$^{-1}$) is in agreement with the {\it Fermi} 
measurement. Direct measurements of the extragalactic UV background
are hampered by the strong dust-scattered Galactic radiation \cite{bowyer91}.
The agreement between the intensity of the UV
background as measured with {\it Fermi}  and that due
to galaxies individually resolved by the Hubble Space Telescope \cite{gardner00}
(3$\pm1$\,nW m$^{-2}$ sr$^{-1}$ versus 2.9--3.9\,nW m$^{-2}$ sr$^{-1}$, 
respectively) shows that the room for any residual diffuse UV emission is small.
This conclusion is reinforced by the good agreement of the {\it Fermi}
measurement and the estimate of the average UV background, at z$\geq$1.7, of
2.2--4.0\,nW m$^{-2}$ sr$^{-1}$ using the proximity effect in quasar spectra \cite{scott00}.

Zero-metallicity population-III stars or low-metallicity
population-II stars  are thought to be the first
stars to form in the Universe and formally marked the end
of the dark ages when, with their  UV light, these objects
started ionizing the intergalactic medium \cite{bromm04}. These stars, whose mass
might have  exceeded  one hundred times the mass of our Sun, are also believed
to be responsible for creating the first metals and dispersing
them in the intergalactic medium \cite{ostriker96,greif07,wise08}. 
A very large contribution of population-III stars to the near-infrared EBL
had already been excluded by \cite{aharonian06}.
Our measurement constrains, according to \cite{raue09,gilmore11},
the redshift of maximum formation of
low-metallicity stars to be at z$\geq$10 and its peak co-moving
star-formation rate to be 
lower than 0.5\,M$_{\odot}$ Mpc$^{-3}$ yr$^{-1}$.
 This upper limit
is already of the same order of the peak star-formation rate of 0.2--0.6\,M$_{\odot}$ Mpc$^{-3}$ yr$^{-1}$ proposed by \cite{bromm02} and suggests
that the peak star-formation rate might be much
lower as proposed by  \cite{tornatore07}.

\clearpage

\bibliographystyle{Science}

\begin{myindentpar}{0.75cm}
Marco Ajello acknowledges generous support from the Fermi guest investigator
program (proposals ID 31117 and 51258) and the Swift and the GROND
teams for observing $\sim$hundred Fermi blazars in an effort to constrain 
their redshifts.
The \textit{Fermi} LAT Collaboration acknowledges generous ongoing support
from a number of agencies and institutes that have supported both the
development and the operation of the LAT as well as scientific data analysis.
These include the National Aeronautics and Space Administration and the
Department of Energy in the United States, the Commissariat \`a l'Energie Atomique
and the Centre National de la Recherche Scientifique / Institut National de Physique Nucl\'eaire et de Physique des Particules in France, the Agenzia 
Spaziale Italiana and the Istituto Nazionale di Fisica Nucleare in Italy, 
the Ministry of Education, Culture, Sports, Science and Technology (MEXT), 
High Energy Accelerator Research Organization (KEK) and Japan Aerospace 
Exploration Agency (JAXA) in Japan, and the K.~A.~Wallenberg Foundation, 
the Swedish Research Council and the Swedish National Space Board in Sweden.
Additional support for science analysis during the operations phase 
is gratefully acknowledged from the Istituto Nazionale di Astrofisica in 
Italy and the Centre National d'\'Etudes Spatiales in France.
\end{myindentpar}

\clearpage

\renewcommand{\thefigure}{S\arabic{figure}}
\renewcommand{\thetable}{S\arabic{table}}

\setcounter{figure}{0}
\setcounter{table}{0}

\section*{Supplements}
%
%
%
\subsection*{Source Selection}

The second LAT AGN Catalog \cite{2LAC} contains
many flat-spectrum radio quasars (FSRQs) and BL Lacertae (BL Lac) objects
whose spectra extend significantly above 10\,GeV.
The classification relies on the conventional definition of BL Lac objects outlined in \cite{stocke91,urry95,marcha96}
 in which the equivalent width of the strongest optical emission line is 
$<$5\,\AA\, and the optical spectrum shows a Ca II H/K break ratio C$<$0.4. 
Sources are then classified also according to the position of the peak
of the synchrotron component  as
low-synchrotron-peaked (LSP, $\nu_{peak}$$<$$10^{14}$\,Hz),
intermediate-synchrotron-peaked (ISP, $10^{14}$$<$$\nu_{peak}$$<$$10^{15}$\,Hz),
 and high-synchrotron-peaked (HSP, $\nu_{peak}$$>$$10^{15}$\,Hz).

\begin{figure*}[ht!]
  \begin{center}
  \begin{tabular}{c}
  	 \includegraphics[scale=0.80]{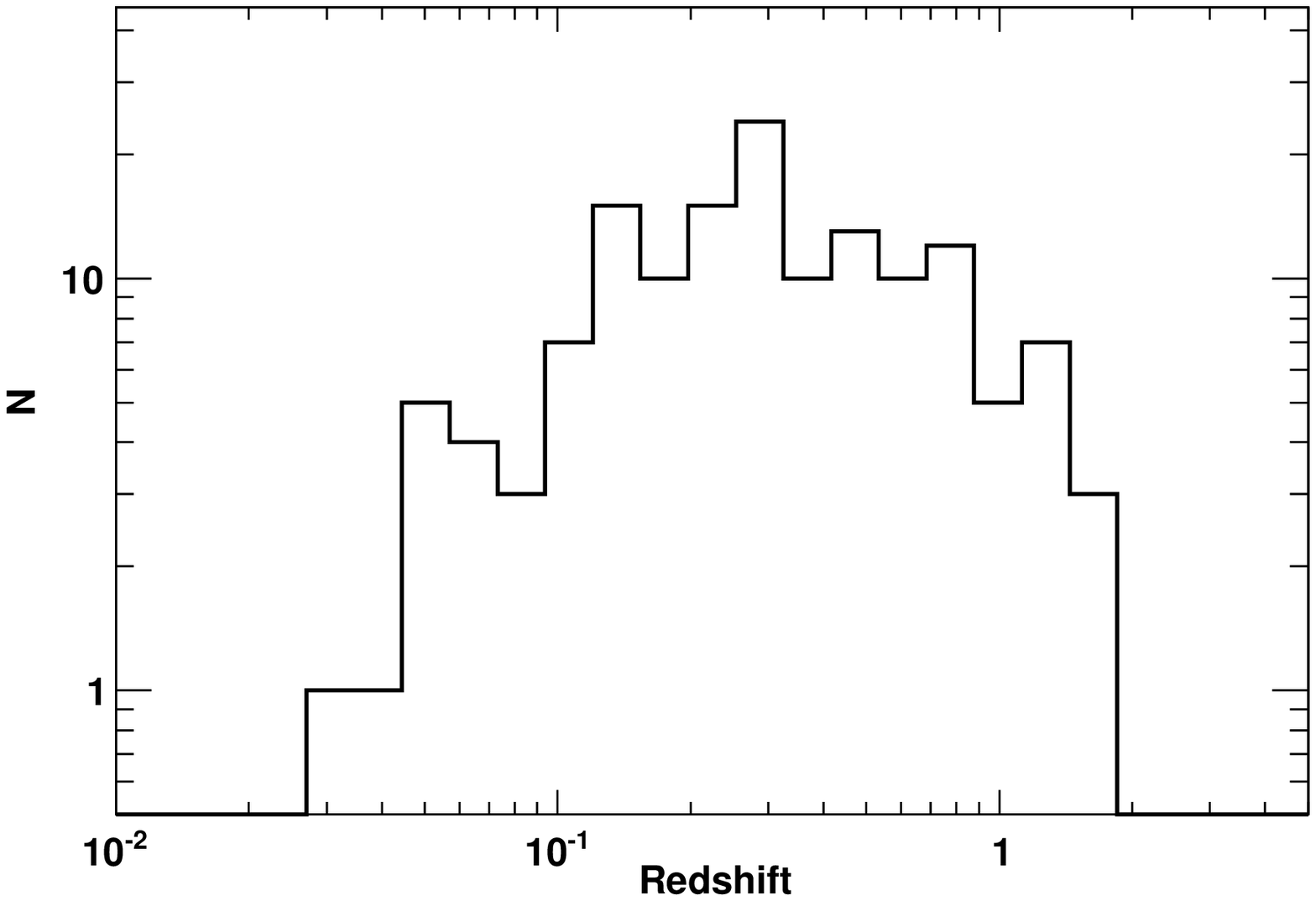} \\
\end{tabular}
  \end{center}
  \caption{Redshift distribution of the BL Lac objects used in this
analysis.
\label{fig:zdistr}}
\end{figure*}

FSRQs generally have soft spectra (photon index greater than 2.3) at GeV
energies, making it more difficult to detect absorption features at energies greater than 10 GeV. Additionally their  spectra might suffer from intrinsic absorption of gamma rays by the photons of the broad line region or of the accretion disk \cite{reimer07}.
This makes them non-ideal candidates to constrain the EBL. We therefore focused on the BL Lac objects. For computational reasons, only 50 sources can be analyzed in one combined likelihood fit. In order to perform our analysis in three redshift intervals we selected the 150 BL Lacs that show the largest detection 
significance in the 3--10\,GeV energy band (the significance is always larger than 3.5\,$\sigma$) with no selection on spectral shape.
Most of the sources (100 out of 150) 
were detected with a significance $\geq$3\,$\sigma$ above 10\,GeV already
after two years of observations \cite{2FGL}.
In the second LAT AGN Catalog catalog \cite{2LAC} only $\sim$190 BL Lac have a redshift measurement,
so our sample is a representative set of {\it Fermi}-detected BL Lacs.
Figure~\ref{fig:zdistr} shows the redshift distribution of the BL Lac
objects employed in this analysis.

%
%
%
\subsection*{Data Analysis}
\label{sec:analysis}

Each source in our sample is analyzed using
 46 months\footnote{From August 4th 2008 to June 1st 2012}
of {\it Fermi}  observations using version v9r27 of the Science
Tools\footnote{http://fermi.gsfc.nasa.gov/ssc/data/analysis/software/}.
 The data were filtered, removing time periods in which the instrument was not
in sky-survey mode, and removing photons whose zenith angle is
larger than 100$^{\circ}$ to limit contamination from the Earth limb emission.  We consider only photons collected within 15$^{\circ}$
of the source position with 1$\leq$E$\leq500$\,GeV. We employ
the P7SOURCE\_V6\footnote{Our analysis is robust against change
of the dataset and IRF.} instrumental response function (IRF) and perform a binned
likelihood analysis. The Galactic and isotropic diffuse 
emissions are modeled using
respectively the gal\_2yearp7v6\_v0.fits and iso\_p7v6source.txt templates.

We rely on the \emph{Composite Likelihood} tool of the \emph{Fermi} software
 to perform  the likelihood maximization. 
This allows us to fit simultaneously  the data from different ROIs with the
aim of constraining the scaling parameter $b$ that is applied to the opacity
curves predicted by each one of the EBL models we tested.
The total number of free parameters across all ROIs in one redshift bin is approximately 1000. It is computationally unfeasible to fit all of these parameters simultaneously. We therefore proceeded in three steps. 
First we fit each ROI individually.
The fit is performed on the entire energy band (1--500\,GeV).
The parameters (i.e. flux and photon index) of all the sources within
4$^{\circ}$ of the target source, along with the parameters of the diffuse
components, are left free to vary.
 More distant sources have parameters frozen at the
values measured in the second {\it Fermi} LAT catalog \cite{2FGL}, 
unless the inspection of the residual map
showed that a given source underwent strong variability.
In that case, the normalizations of those sources were left free
in the fit.
The spectra of the sources of interest are modeled using 
a {\it LogParabola}  model \cite{2FGL}. 
We then proceed to re-optimize the parameters of the
source of interest in each ROI up to that energy for which the EBL absorption
becomes no longer negligible (see the definition of $E_{crit}$ in the main text). 
In the third step we  fix the curvature of the {\it LogParabola}  model to
the  value obtained in the previous fit.
Finally, we use the {\it Composite Likelihood} to constrain
the gamma-ray opacity. The {\it Composite Likelihood} allows the user
to tie any parameters between  any ROIs. 
In our case the only tied parameter is the renormalization 
factor $b$ of the opacity of a given EBL model.

The significance of our finding can be evaluated using the Test Statistic (TS)
as $TS=2(log \mathscr{L}(b) -log \mathscr{L}(b=0)) $, where $\mathscr{L}$
is the likelihood function and  the $b=0$ case
(i.e. no EBL absorption) represents the null hypothesis.
Since the null hypothesis is a special case of the hypothesis we
test, we expect the $TS$ to be distributed as a $\chi^2$ with one degree
of freedom (see also next sections). 
This allows one to transform the TS into the corresponding
number of standard deviations of a Gaussian distribution as $n_{\sigma}=\sqrt{TS}$.

%
%
%
\subsection*{Modeling the Intrinsic Blazar Spectrum}
\begin{figure*}[ht!]
  \begin{center}
  \begin{tabular}{c}
  	 \includegraphics[scale=0.50]{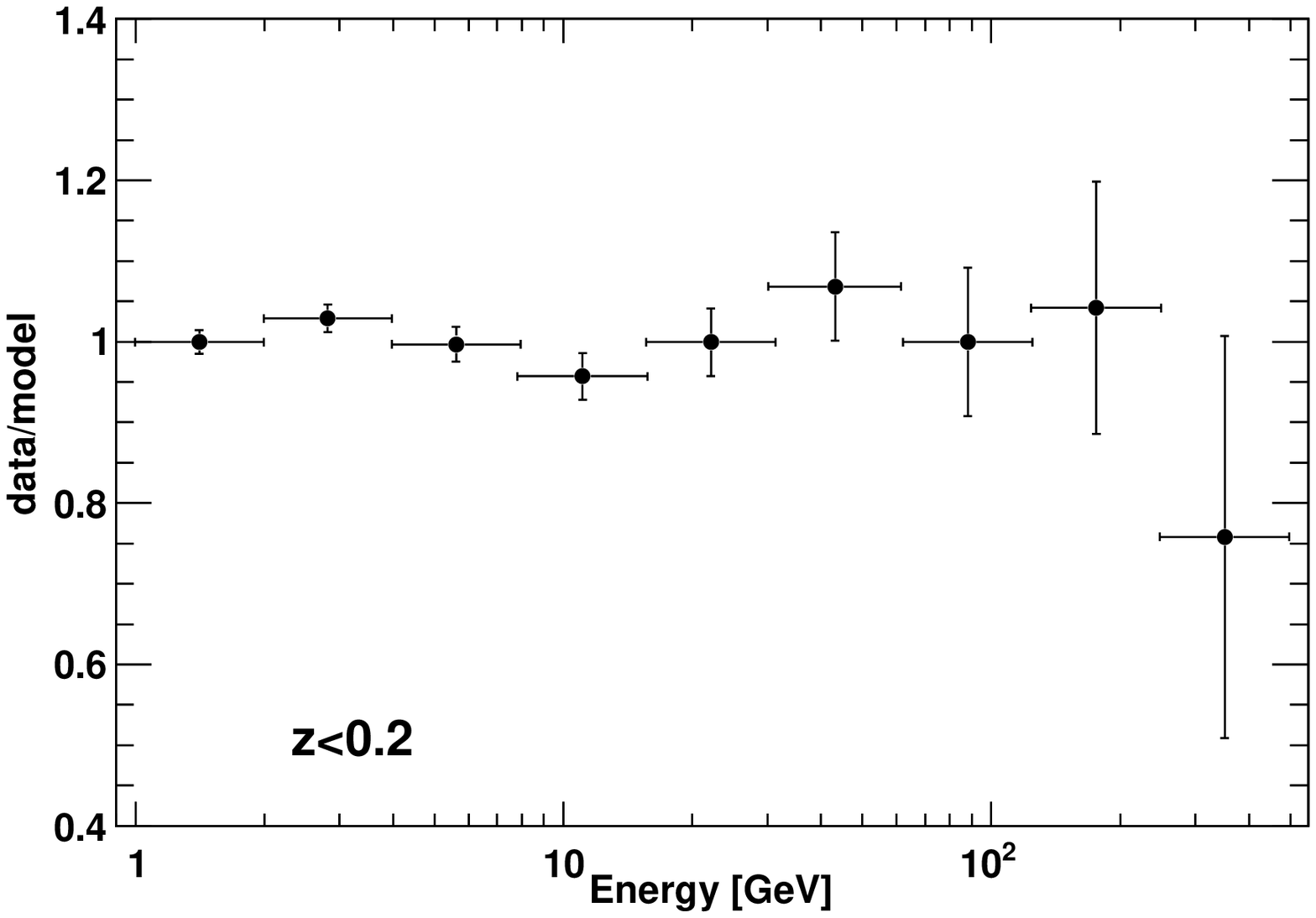} \\
\end{tabular}
  \end{center}
  \caption{Average of the residuals with respect to the best fit
models for all the blazars in the z$<$0.2 redshift bin.
\label{fig:residuals_lowz}}
\end{figure*}

One of the basic assumptions of this analysis is that on average
 the spectrum of a BL Lac blazar
can be adequately modeled using a {\it LogParabola}
in the 1-500\,GeV band. Figure~\ref{fig:residuals_lowz} shows
the residuals of the best fits to all the sources in the z$<$0.2
interval. It is apparent that the {\it LogParabola} provides
a good representation of the spectra of blazars in our sample.

In an additional test we artificially decreased
the critical energy $E_{crit}$ for all the sources in the z$<$0.2 interval
from the typical $\geq$120\,GeV (for z$<$0.2) to $\sim$40\,GeV which
is representative of the z$\geq$0.5 case. Even in this case the results
are unchanged, showing that the properties of the intrinsic
spectrum can be determined over a more restricted energy range and
can be safely extrapolated above the critical energy. 

Blazars are however known to have complex spectra that cannot
always be modeled using simple empirical functions. To further
test our basic assumption, we fit the spectra of bright blazars
with coverage in the GeV and TeV band. We rely on the published
spectra of Mrk 421 \cite{lat_421}, Mrk 501\cite{lat_501} and RBS 0413\cite{rbs0413}. We find that in the energy range relevant for the present analysis
the {\it LogParabola} model provides an adequate fit
to  all these spectra. A further confirmation comes from our validation
studies using simulations of broad-band blazar spectra (see next sections).
We thus conclude that in a small energy range like the one
adopted here (1--500\,GeV), the {\it LogParabola} is an adequate
representation of the intrinsic blazar spectrum.

%
%
%
\subsection*{Results}

Table~\ref{tab:results} reports the results of the joint-likelihood
fit for all the predictions of the EBL models we tested. 
In many cases, a given EBL model (because of tunable parameters or
uncertainties connected to e.g, galaxy evolution, star-formation rate etc.)
comprises several different predictions for the opacity. In this
case, we use these prediction at face value and test them against
our observations. 
All the models with the exclusion
of \cite{kneiske04}\footnote{The model of 
 \cite{kneiske04} adopts a value of H$_0$=65\,km s$^{-1}$ Mpc$^{-1}$.
We checked that neglecting the change of H$_0$ for the model of 
 \cite{kneiske04}  introduces an error on
the opacity of $<$7\,\% which has negligible influence on our analysis..
} rely on a standard concordance cosmology 
(H$_0$=70\,km s$^{-1}$ Mpc$^{-1}$, $\Omega_M$=1-$\Omega_{\Lambda}$=0.3).
For each model we report the significance of the absorption feature
detected once we allow the opacity to be renormalized by $b$.
The significances have been
obtained as a square root of the  sum  of the TS in the 3 redshift intervals
while the $b$ parameter as reported in Table~\ref{tab:results} represents the 
weighted
average\footnote{The weights are 1/$\sigma_b^2$ where $\sigma_b$ is
the uncertainty on the $b$ parameters.} of the 3 $b$ values determined in the 3 redshift bins.  For example, for the model of \cite{franceschini08}
we measure: 
$b$=1.18$^{+0.94}_{-0.81}$ (TS$\approx$4 for z$<$0.2),
$b$=0.82$^{+0.41}_{-0.30}$ (TS$\approx$7 for 0.2$\leq$z$<$0.5), and
$b$=1.29$^{+0.43}_{-0.36}$ (TS$\approx$25 for 0.5$\leq$z$<$1.6).
The weighted average yields (as reported in Table \ref{tab:results})
 $b=1.02\pm0.23$ and TS$\approx$36.

We also quantify the level of compatibility of the prediction of 
a given model with the {\it Fermi} data, by comparing
the likelihood of the original model ($b=1$) to the one of the best fit scenario
($b$ free to vary). For $b$ values significantly different than 1 a given EBL
model predicts an attenuation larger than observed.
Table~\ref{tab:results} shows that the intensity of the
optical-UV background in 4 of the 15 models we tested
is not compatible with the {\it Fermi} data at $\geq3$\,$\sigma$.

\begin{figure*}[ht!]
  \begin{center}
  \begin{tabular}{c}
  	 \includegraphics[scale=0.83,clip=true,trim=0 0 0 0]{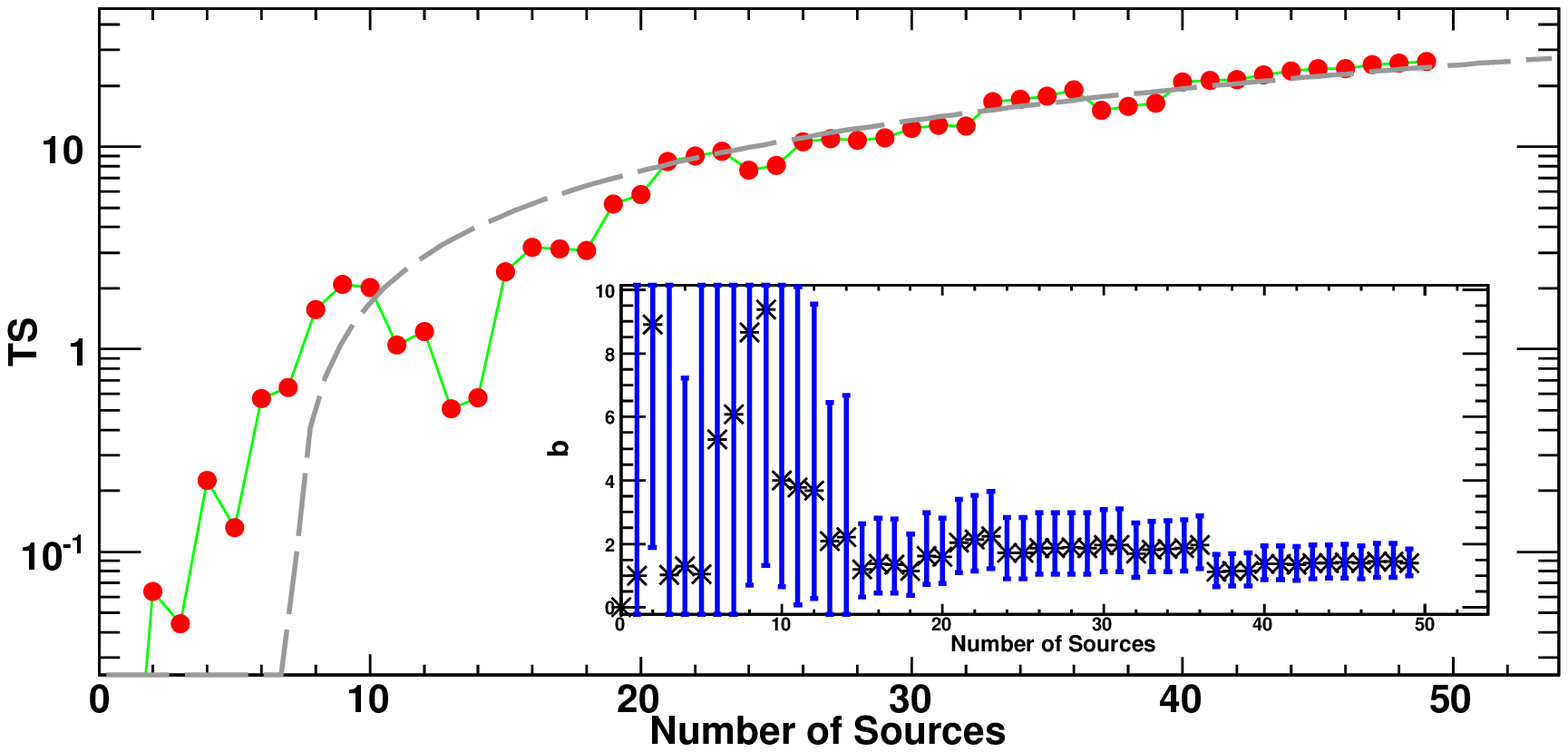} \\
\end{tabular}
  \end{center}
  \caption{Increase in the TS value of the (renormalized) EBL model
of \cite{franceschini08} produced in the joint-likelihood fit 
(to the 0.5$\leq$z$<$1.6 interval) while
adding one source at a time. The sources have been sorted in redshift 
(from lowest to highest). The dashed line shows the best-fit linear increase
of the TS with the number of sources. The inset shows the best-fit value
of the renormalization parameter $b$ applied to the opacity predicted
by \cite{franceschini08} (see text for details).
\label{fig:tsincrease}}
\end{figure*}

Figure~\ref{fig:tsincrease} shows that the significance of the absorption
feature is detected collectively\footnote{In Figure~\ref{fig:tsincrease}
the sources have been sorted in redshift from low to high redshift.} among the BL Lacs considered and is not attributable to just a few sources.
Indeed, it is apparent that the TS increases linearly with the number of sources
as expected in a background-limited scenario.

As shown already in Figure~\ref{fig:tau}, our analysis probes absorption
signatures that are a factor $\sim$4 times weaker than those probed
in \cite{lat_ebl10} which relied on only one year of data.
 Even a conservative 95\,\% upper limit derived
from this analysis ($b_{UL}=1.5$ referred to the model of \cite{franceschini08},
i.e. $\tau_{\gamma \gamma}$$<$0.6 at z$\approx$1 and E=77\,GeV)
is a factor $\sim$2 below the one reported by \cite{lat_ebl10}.

Our analysis relies on the assumption
that BL Lac spectra (and HSP in particular) do not change dramatically
across the z=0.2 redshift barrier. In the next section
we address possible sources of systematic effects.

\begin{deluxetable}{lccc||c}
\tablewidth{0pt}
\rotate
\tablecaption{Joint-likelihood results for different EBL models.
\label{tab:results}}
\tablehead{\colhead{Model\tablenotemark{a}} & \colhead{Ref.\tablenotemark{b}} &
\colhead{Significance of $b$=0 Rejection\tablenotemark{c}} &
\colhead{$b$\tablenotemark{d}      }          & 
\colhead{Significance of $b$=1 Rejection\tablenotemark{e}}
}
\startdata

{\it Stecker et al. (2006) -- fast evolution} &\cite{stecker06} & 4.6  & 0.10$\pm0.02$ & 17.1\\ 
{\it Stecker et al. (2006) -- baseline} &\cite{stecker06} & 4.6  & 0.12$\pm0.03$ & 15.1\\ 
{\it Kneiske et al. (2004) -- high UV} &\cite{kneiske04} & 5.1  & 0.37$\pm0.08$ & 5.9\\ 
{\it Kneiske et al. (2004) -- best fit}&\cite{kneiske04} & 5.8  & 0.53$\pm0.12$ & 3.2\\ 
{\it Gilmore et al. (2012) -- fiducial}&\cite{gilmore12} & 5.6  & 0.67$\pm0.14$ & 1.9\\ 
{\it Primack et al. (2005)} &\cite{primack05} & 5.5  & 0.77$\pm0.15$ & 1.2\\ 
{\it Dominguez et al. (2011)}&\cite{dominguez11} & 5.9  & 1.02$\pm0.23$ & 1.1\\ 
{\it Finke et al. (2010) --  model C} &\cite{finke10}& 5.8  & 0.86$\pm0.23$ & 1.0\\ 
{\it Franceschini et al. (2008)} &\cite{franceschini08}& 5.9  & 1.02$\pm0.23$ & 0.9\\ 
{\it Gilmore et al. (2012) -- fixed} &\cite{gilmore12}& 5.8  & 1.02$\pm0.22$ & 0.7\\ 
{\it Kneiske \& Dole (2010)} &\cite{kneiske10}& 5.7  & 0.90$\pm0.19$ & 0.6\\ 
{\it Gilmore et al. (2009) -- fiducial} &\cite{gilmore09}& 5.8  & 0.99$\pm 0.22$ & 0.6\\

\enddata
\tablenotetext{a}{Models tested are 
implemented in the {\it Fermi} Science Tools. 
As an example the recent model of \cite{stecker12} which is not
defined for E$\geq$250$/(1+z)$\,GeV could not be used, but its predictions
are for completeness reported in Figure~\ref{fig:tau}.}

\tablenotetext{b}{Reference number in the `References and Notes' section
of the main text.}

\tablenotetext{c}{Significance, in units of $\sigma$, of the 
attenuation in the spectra of blazars when a given
EBL model is scaled by the factor $b$. In this
case $b$=0 (i.e. no EBL absorption) constitutes the null hypothesis.}
\tablenotetext{d}{This column lists the maximum likelihood values 
and 1\,$\sigma$ confidence ranges for the opacity scaling factor.}
\tablenotetext{e}{Here the $b$=1 case (i.e. EBL absorption as predicted 
by a given EBL model) constitutes the null hypothesis. This column shows
the compatibility (expressed in units of $\sigma$)
of the  predictions of EBL models with the {\it Fermi}
observations. Large values mean less likely to be compatible.}

\end{deluxetable}

%
%
%
\subsection*{Validation of the Analysis using Simulations}

We validated our analysis using simulated datasets starting
from physically motivated spectral energy distributions (SEDs)
of BL Lacs. We rely on SEDs that reproduce well the  range
of peak curvatures, peak frequencies (for both the synchrotron
and high-energy component), 
and gamma-ray photon indices observed for the LAT-detected
BL Lacs. These SEDs have been produced  using the same numerical code presented in 
\cite{tramacere11} and    take into account  all the important
effects  that contribute to determine the intrinsic curvature of the gamma-ray spectrum of
the {\it Fermi} blazars. Of particular importance are
those effects that depend on the curvature of the 
electron distribution, as well as those related to the Thomson (TH) to Klein-Nishina (KN) 
transition, in the inverse  Compton process (IC). The latter effect is crucial
since  it is well  known \cite{massaro06,tramacere11} that the TH/KN transition 
in the  IC cross section
can result in a steepening of the high-energy branch of the IC SED, compared to the low-energy one. Our simulations fully take into account this effect, allowing us
to study the potential bias in the EBL estimate.

We performed $\sim$500 simulations randomly selecting 50 SEDs from our template
library. To each SED we assign a redshift (in the 0.5$\leq$z$<$1.6 range)
and a 0.1--100\,GeV flux 
randomly extracted from those of BL Lacs in the second LAT AGN Catalog \cite{2LAC}. We then proceeded to simulate the data expected from those sources
based on {\it Fermi}'s instrument response function and pointing history.
Each one of those realizations was processed with the same analysis
chain used on actual data.

\begin{figure*}[ht!]
  \begin{center}
  \begin{tabular}{c}
  	 \includegraphics[scale=0.85]{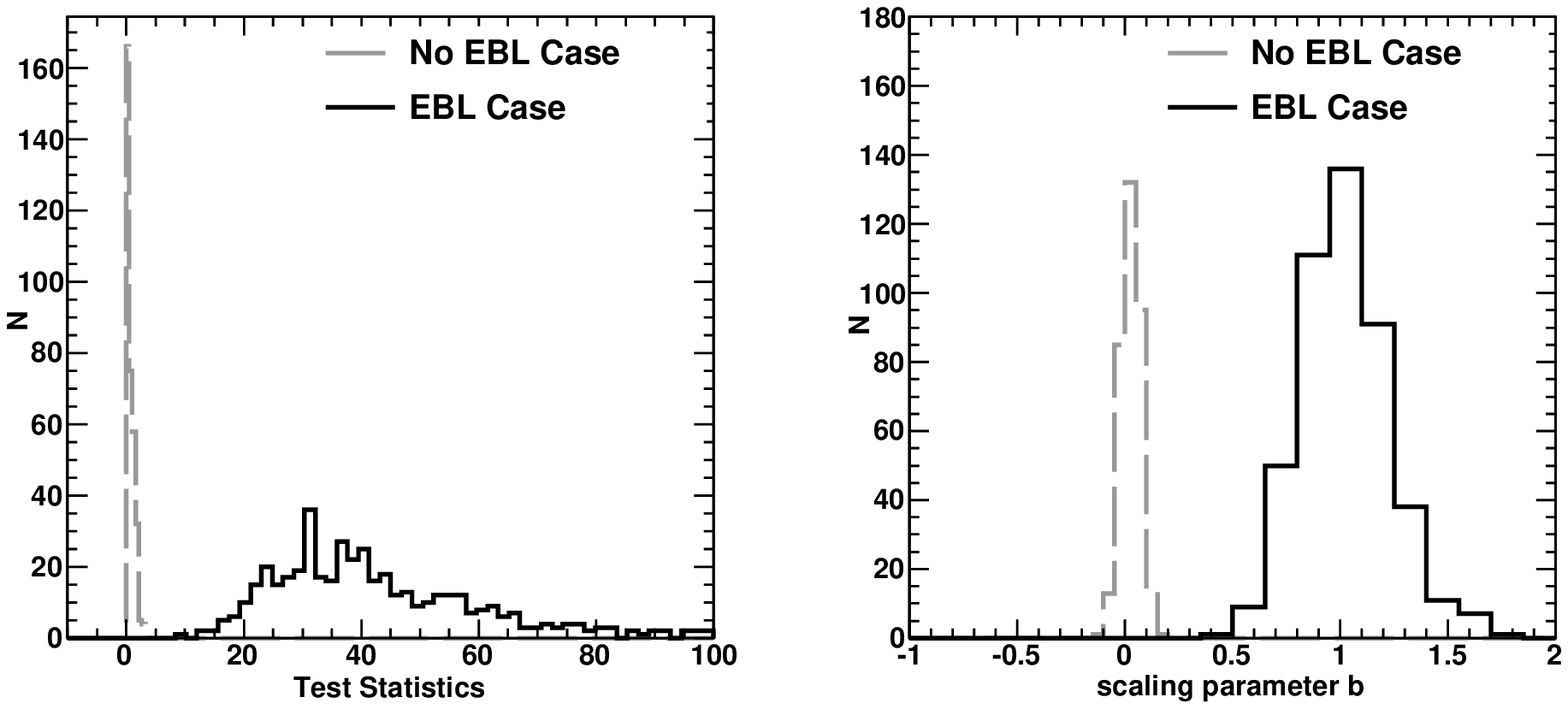} \\
\end{tabular}
  \end{center}
  \caption{Results from joint-likelihood fits to the simulated datasets. The left panel
shows the TS of the detection of the attenuation produced
by the EBL for the case in which no EBL was applied to the SEDs (dashed line)
and the case for which an attenuation consistent with the one
of \cite{franceschini08} was used (solid line). The right panel shows
the best-fit $b$ parameter for the same two scenarios.
\label{fig:sim}}
\end{figure*}

Figure~\ref{fig:sim} presents the distribution of TS (left panel)
and best-fit $b$ parameters (right panel) for the simulated blazar
population under the following two scenarios. First,
no EBL attenuation was included, in order to check whether our
analysis would pick a signal consistent with EBL attenuation when
this was not present (i.e. a false positive). The dashed lines in the 
panels show that our analysis yields very small TS values
and $b$ values compatible with zero, as expected for a robust analysis.

In the second scenario, the SED models were attenuated using the
opacities of the model of \cite{franceschini08}. The results
summarized in Figure~\ref{fig:sim} show that TS  
values $\geq$20 and $b$ values compatible with 1 (within
the statistical uncertainty) are correctly retrieved.
Moreover, the simulations show that our analysis is free of any major
systematic uncertainty  since the average values of $b$ retrieved for 
both scenarios are compatible with the expected ones (0 and 1 respectively)
 within 1\,\%.

%
%
%
\subsection*{Additional Tests}
A number of additional tests have been performed to test the stability
of the results presented in the previous sections. These are documented below.

\subsubsection*{Sample Selection}
The selection criteria used in the previous sections were chosen {\it a priori}
to yield a uniform source population that could be used to  probe
in a sensitive and least-biased way the EBL-induced absorption feature.
Several tests have been performed to check this assumption.

The BL Lac population is known to evolve in redshift with
HSP objects detected predominantly at low redshift and LSPs detected
predominantly at larger redshifts with ISPs bridging the gap\footnote{The sample
used in the previous sections contains for z$<$0.2 35 HSPs, 10 ISPs and 5 LSPs.
The 0.2$\leq$z$<$0.5 sample comprises 27 HSPs, 18 ISPs, 5 LSPs, while the z$\geq$0.5
sample contains 10 HSPs, 19 ISPs, and 21 LSPs.}.
This could artificially induce a decline in the combined (produced by
all subtypes) BL Lac spectrum at an energy that decreases with
increasing redshift, and hence could mimic the effect of EBL-caused
absorption. To test whether such bias affects our results we run the
analysis separately for the three source classes (HSPs, ISPs and LSPs). For each of them we define two redshift bins: z$<$0.2 and z$\geq$0.2.
In the lowest redshift bin, due to the low statistics\footnote{
In the z$<$0.2 bin
there are respectively 41 HSPs, 10 ISPs and 5 LSPs; while in
the z$\geq$0.2 there are 50 HSPs, 30 ISPs, and 23 LSPs.} the joint likelihood
yields no significant detection of EBL softening. For the HSPs in the low
redshift bin, the joint likelihood yields 
$b$=1.46$^{+0.89}_{-0.78}$ and a TS=4.1. 
At higher redshift, the results are:
$b=$1.04$^{+0.32}_{-0.28}$ (TS=24), 
$b=$2.13$^{+1.11}_{-0.98}$ (TS=9), and
$b=$0.43$^{+0.67}_{-0.71}$ (TS=1),
for HSPs, ISPs and LSPs respectively.
The weighted average of the above values yields $b=$1.04$^{+0.27}_{-0.24}$
and a total TS of $\sim$38. These results are in very good agreement
with the ones presented in the previous sections and show that high-redshift
LSPs are not responsible for the observed spectral softening, and HSPs
are mainly responsible for it.
Moreover, since the signal is dominated by only one source
population (namely HSPs) the probability that selection effects from
changing the sample composition with redshift strongly affect our result
is rather low. 

We also checked if the highest redshift HSPs are driving the result.
To this extent we isolated (from the above sample) the 6 HSPs with
the highest redshift. The joint likelihood yields a TS=6 and 
$b=$1.99$^{+1.83}_{-1.10}$ indicating that the detection of the EBL cut-off
is not caused by a few high redshift sources.

We also tested the population of BL Lacs that did not pass our
selection criteria. This comprises 32 objects detected with 
a rather low significance in the 3-10\,GeV over a 
redshift range 0.03-1.9 (average of 0.4). The joint likelihood
yields $b$=1.93$^{+2.73}_{-1.39}$ and TS=3.5 compatible with our result.
Thus, including low significance sources marginally 
improves the results of our analysis.

\subsubsection*{Binning in Redshift}
We bin the sources in redshift to cope with
the maximum number of free parameters (i.e. 100)
that can be optimized in a single maximum likelihood fit.
Our analysis is however independent of the redshift binning used.
This is already evident from the tests reported above using separately
HSPs, ISPs and LSPs and a different redshift binning (with respect to
the one used in the main analysis). However, we performed an additional
test and divided the most constraining redshift bin (i.e. 0.5$\leq$z$<$1.6)
into  halves (at z$\approx$0.75) containing 25 sources each.
When fitting the {\it Fermi} data using the model of \cite{franceschini08},
we obtain $b$=1.71$^{+1.12}_{-0.82}$ (TS=10) and $b$=1.17$^{+0.47}_{-0.39}$ (TS=15.4) for the 0.5$\leq$z$<$0.75 and 0.75$\leq$z$<$1.6 bins respectively.
The weighted average of the above values yields
$b$=1.26$^{+0.43}_{-0.35}$ and the total TS=25.4.
This can be compared to $b$=1.29$^{+0.43}_{-0.36}$ and  TS=25.1
obtained for the full 0.5$\leq$z$<$1.6 redshift bin. Our analysis
is thus robust against choices of the redshift bins.

\subsubsection*{Source Variability}
Blazars are inherently variable objects with variability
in flux of up to a factor 10 or more. Throughout  this work
we use spectra of blazars accumulated over 46\,months of observations
and blazar variability might in principle bias our result.
In the {\it Fermi} sample, the most variable BL Lac objects are those
belonging to the LSP and ISP class and it has been shown already that the entire
analysis can be confirmed without using those two classes.

To test possible biases deriving from variability we selected the
30 most variable, according to \cite{2FGL,2LAC}, and most
significant (in the 3--10\,GeV band) BL Lac objects. Repeating
the entire analysis using this sample, we obtain $b=1.20^{+0.62}_{-0.52}$
and a TS=9. These results are in good agreement with
the ones presented in the previous section and show that
blazar variability does not affect our results.

\subsubsection*{Critical Energy}
Our measurement relies on the extrapolation of the intrinsic (i.e. unabsorbed)
spectrum above the critical energy $E_{crit}$. Aggressive choices of
this energy (i.e. large values) might bias the result towards a low level
of opacity. We show that this is not the case for our measurement and
that our analysis is robust against conservative choices of $E_{crit}$.
In order to demonstrate it, we imposed $E_{crit}$=10\,GeV for any source
in the most constraining redshift interval: i.e. 0.5$\leq$z$<$1.6.
For most EBL models, with the exception of those presented
 in \cite{stecker06}, 
the opacity is negligible at 10\,GeV up to z$\approx$1.6.
When fitting the {\it Fermi} data using the model of \cite{franceschini08}, 
our analysis yields a value of $b$=1.24$^{+0.44}_{-0.37}$ and  TS$\approx$22
for the constant  $E_{crit}$=10\,GeV case. This can be compared to 
$b$=1.29$^{+0.43}_{-0.36}$ and  TS$\approx$25.1 for the variable $E_{crit}$
case presented in the main text. We thus conclude that our analysis is
robust against any conservative choice of $E_{crit}$.

\subsubsection*{Systematic Uncertainties of the Instrument Response Function}
In order to gauge the systematic uncertainties of this analysis
we use the IRF bracketing method as described in \cite{lat_perf}.
By deriving two different sets of IRFs  and repeating the entire analysis
 we find that the systematic uncertainty on 
the opacity $\tau_{\gamma \gamma}$ is of the order $\sim$7\,\%.

Moreover, the entire analysis has been repeated using the P7CLEAN\_V6 photon 
selection and IRF. The results presented in the above sections
are fully confirmed and the systematic 
uncertainty on the opacity $\tau_{\gamma \gamma}$ due to changing
event selection and IRF is of the order 3\,\%. Thus, we consider the 
total systematic uncertainty on $\tau_{\gamma \gamma}$ to be $\sim$10\,\%.

%
%
%
\subsubsection*{Source Intrinsic Effects}

The blazar evolution with luminosity as described by 
the blazar sequence \cite{ghisellini98,fossati99,ghisellini09b,meyer11}
and the change of blazar type with redshift might in principle
produce an intrinsic cut-off that changes with redshift.
An intrinsic spectral cut-off might be 
detected if the part of the blazar spectrum sampled by {\it Fermi}
corresponds to the tail of the electron distribution.
In our case, this might happen only for the LSP sources and
it has already been shown (see previous section)
 that their contribution to the total signal is negligible.
On the other hand most of the signal is dominated by HSPs that
are detected by {\it Fermi} right below or
at the peak of the inverse Compton component,
thus excluding the possibility that their emission is produced in the tail of 
the electron distribution.

However, in order to study the compatibility of the detected signal
with an intrinsic origin, we assumed that all blazars have an
exponential cut-off at a source-frame energy $E_0$. We also assumed that
the blazar spectrum moves to lower frequency for increasing source
luminosity as dictated by the so-called blazar sequence \cite{ghisellini98,fossati99,ghisellini09b,meyer11}. This might represent the case in which 
the maximum electron energy depends on  the jet power or the luminosity.
Figure~\ref{fig:residuals} shows that this model ($E_0$ is fitted to the data)
provides a poor representation of the {\it Fermi} data.
Even assuming that the maximum electron energy depends on a
variable power of the luminosity does not improve the results.
The result is even worse if the blazar sequence is not invoked.
We thus conclude that the impact of source intrinsic effects on
our analysis is likely minor.

%
%
%
\subsubsection*{Reprocessed Emission}

The electron-positron pairs generated in the interaction of  gamma rays
and  EBL photons can initiate a cascade by subsequent Compton scattering
of the photons of the Cosmic Microwave Background. Typically the cascades
originated by TeV gamma rays are reprocessed in the GeV energy range.
 In the case of a
 weak intergalactic magnetic field (IGMF), the pairs are not deflected out of the
line of sight \cite{plaga95} and the reprocessed emission can be a substantial
fraction of (or even dominate) the total source signal in the $\lesssim$100\,GeV band.
The intensity of the reprocessed emission depends primarily on the EBL density,
on the intensity of the IGMF, on its coherent length 
and on the position of the peak of the high-energy component of a blazar
\cite{tavecchio11}. For typical coherent lengths of $\sim$1\,Mpc and SED peaks
located at or below 10\,TeV, lower  limits of B$\gtrsim$10$^{-15}$\,G have
been inferred \cite{neronov10,tavecchio11}. For these intensities of the IGMF
the reprocessed emission is expected to be subdominant with respect to the
intrinsic component \cite{vovk12}. While it is true that these estimates
are based on an implied activity of the source  of a few million years \cite{dermer11},
the majority of the {\it Fermi} BL Lacs is expected to have the peak
of the high-energy component
located at $\leq$1\,TeV \cite{tavecchio10} greatly reducing 
the amount of reprocessed emission in the GeV band \cite{tavecchio11}.
Similar considerations (i.e. suppression of the reprocessed emission
by the IGMF) hold also for the case in which blazars are also sources
of ultrahigh energy cosmic rays \cite{mannheim93,waxman96,essey10}.


\begin{thebibliography}{10}
\bibitem{note1}
We define the critical energy $E_{crit}$ such that
less than 5\,\% of the source photons are absorbed for 
the EBL model of \cite{franceschini08} below this energy.
\end{thebibliography}
\end{document}